\documentstyle[prb,aps,epsf,floats]{revtex}

\begin{document}
\draft

\twocolumn[\hsize\textwidth\columnwidth\hsize\csname@twocolumnfalse%
\endcsname

\title{
Distributions of gaps and end-to-end correlations in random 
transverse-field 
Ising spin chains  
}

\author{Daniel. S. Fisher}
\address{Physics Department, Harvard University, Cambridge, MA 02138 }
\author{A. P. Young}
\address{Department of Physics, University of California, Santa Cruz, 
CA 95064}

\date{\today}

\maketitle

\begin{abstract}
A previously introduced real space renormalization-group treatment of the 
random 
transverse-field
Ising spin chain is extended to provide detailed
information on the distribution of the energy gap and the end-to-end
correlation function for long chains with free boundary conditions. Numerical 
data,
using the mapping of the problem to free fermions, are found to be in good
agreement with the analytic finite size scaling predictions. 

\end{abstract}

\vskip 0.3 truein
]

\section{Introduction}

It has been known for some time that the properties of many random systems are 
dominated by rare regions.  In random quantum systems at low or zero 
temperature, this effect is particularly pronounced.  For example, in random 
quantum magnets, {\it typical} spin correlations at zero temperature can decay 
rapidly with distance while {\it average} correlations - which are measured by 
scattering experiments - decay much more slowly.  In extreme cases, the  
average 
correlations and related quantities such as susceptibilities are dominated by 
rare pairs of spins whose correlations can be large even when they are very far 
apart. 

In the last few years, it has become clear that the simplest of all random 
quantum systems that undergoes a phase transition, the random transverse field 
Ising spin chain, exhibits these phenomena in a particularly dramatic manner.
Surprisingly, a tremendous amount of {\it analytic} information can be obtained 
about this system in the low energy, long length scale regime near the critical 
point from a real space renormalization group (RG) analysis that yields exact 
asymptotic results \cite{dsf}.  So far, most of the results have concerned the 
behavior of average quantities, such as the mean spin-spin correlation function 
and the susceptibility. But the general structure of the RG implies that 
typical 
correlations are very broadly distributed and take an entirely different form 
than the average correlations.  Numerical studies \cite{yr} have yielded the 
distributions of spin correlations and energy gaps and found scaling behavior 
consistent with the general expected structure but there have, so far, not been 
analytic predictions for the distribution functions.

In this paper, we focus on the behavior of random transverse field Ising chains 
with free boundary conditions and study the scaling behavior of the 
distributions of the energy gap and of the correlations between the spins at 
opposite ends of the chain \cite{IR}.  We obtain analytic results for these distributions 
in the long-chain scaling limit and compare them with numerical results on 
finite length chains. The connections between the couplings in each particular 
sample and its properties are also studied.

In addition to its own intrinsic interest, it is hoped that better 
understanding 
of correlations in this seemingly simple system may lead to progress on other 
more complicated systems - particularly higher dimensional random quantum 
systems with phase transitions which are {\it not} governed by RG fixed points 
accessible by conventional field theoretic methods.

\section{The Model}
The model that we study is the
one-dimensional random transverse-field Ising
chain with Hamiltonian
\begin{equation}
{\cal H} = -\sum_{i=1}^{L-1} J_i \sigma^z_i \sigma^z_{i+1} -
\sum_{i=1}^L h_i \sigma^x_i \ .
\label{ham}
\end{equation}
Here the $\{\sigma^\alpha_i\}$ are Pauli spin matrices, and the
interactions $J_i$ and transverse fields $h_i$ are both independent
random variables, with distributions $\pi(J)$ and $\rho(h)$ respectively. Since 
the model is in one dimension, we
can perform a gauge transformation to make all the
$J_i$ and $h_i$ positive.
The numerical work used the following rectangular distributions:
\begin{eqnarray}
\pi(J) & = & 
\left\{
\begin{array}{ll}
1 & \mbox{for $ 0 < J < 1$,} \\
0  & \mbox{otherwise},
\end{array}
\right.
\nonumber \\ 
\rho(h) & = & 
\left\{
\begin{array}{ll}
h_0^{-1} & \mbox{for $ 0 < h < h_0$,} \\
0  & \mbox{otherwise.}
\end{array}
\right.
\label{dist}
\end{eqnarray}
which  are characterized by a
single control parameter, $h_0$. 
The lattice size is $L$, and, in this paper, we will consider  {\em free}
boundary conditions.

Defining
\begin{eqnarray}
\Delta_h & = & \overline{\ln h} \nonumber \\
\Delta_J & = & \overline{\ln J}
\end{eqnarray}
where the overbar denotes an average over disorder,  the
critical point occurs when\cite{sm,dsf}
\begin{equation}
\Delta_h = \Delta_J \ .
\label{crit}
\end{equation}
Clearly this is satisfied if the distributions of bonds and fields are
equal, and the criticality of the model then follows from duality\cite{dsf}.

We define a quantity, $\delta$, which measures the deviation from criticality
by
\begin{equation}
\delta = {\Delta_h - \Delta_J \over \mbox{var}\, h + \mbox{var}\, J}
= {l_v \over 2} \left(\Delta_h - \Delta_J\right)
\label{delta}
\end{equation}
where $l_v$ is given by
\begin{equation}
l_v = {2 \over \mbox{var}\, h + \mbox{var}\, J} ,
\label{lv}
\end{equation}
with ``var'' the variance over the distribution of the couplings in the
Hamiltonian.

For the distribution of Eq.~(\ref{dist}), we see that from Eq.~(\ref{crit}), 
the critical point corresponds to $h_0=1$,
and, from Eq.~(\ref{delta}), $\delta$
is given by
\begin{equation}
\delta = {1 \over 2} \ln h_0 ,
\end{equation}
while Eq.~(\ref{lv}) gives 
\begin{equation}
l_v = 1.
\label{lv1}
\end{equation}

We will focus on the energy gap, $\Delta E$, between the ground state and first
excited state, and the end-to-end correlation function
\begin{equation}
C_{1,L} \equiv  \langle \sigma^z_1 \sigma^z_L \rangle.
\end{equation}
This latter quantity, 
for reasons that are somewhat different than those in the pure Ising model, is 
rather easier to analyze than bulk correlations \cite{IR}.  We first summarize the RG 
method and the basic picture of the system that emerges, then present the 
analytic results, and finally  present the numerical results and compare them 
with the analytic predictions. In the Appendix, the distribution of the energy 
gap with periodic (rather than free) boundary conditions is computed. 

\section{Analytical Results}

In this section  we use the methods of Ref.~\onlinecite{dsf}
to obtain the scaling forms of the distribution of
$\ln \Delta E$ and $\ln C_{1, L}$ which will be
valid in the limit $L$, $-\ln \Delta E$ and $-\ln
C_{1, L}$ large and $|\delta|$, the distance to the
critical point small.  We will follow the notation
of Ref.~\onlinecite{dsf} and refer to equations in that paper by,
e.g. F(5.28).

\subsection{Renormalization Group Analysis}

The basic renormalization group (RG) procedure
from which the low energy, long distance behavior
can be obtained is as follows: First choose the
largest coupling in the system
\begin{equation}
\Omega_I\equiv\mbox{max}\{h_i, J_i\}.
\label{eq:A1}
\end{equation}
If the coupling is a field, say $h_2$, then
decimate the associated spin, leaving, by second
order perturbation theory, an effective bond
\begin{equation}
\tilde{J}_{13}\approx\frac{J_{12}J_{23}}{h_2}<h_2
\label{eq:A2}
\end{equation}
which couples together the spins on the two sides
of the decimated one.  For our purposes here, it
is important to keep track of the correlations, so
we note that first order perturbation theory
yields a correlation of the decimated spin with
each of its neighbors, e.g.:
\begin{equation}
\langle \sigma^z_1\sigma^z_2 \rangle \approx\frac{J_{12}}{H_2}.
\label{eq:A3}
\end{equation}
  
If instead the strongest coupling is a bond, say
$J_{12}$, then join together the spins on both
sides of this bond to make a spin cluster, and
assign it an effective spin cluster variable
$\tilde{\sigma}_{12}.$ The couplings between this
cluster and its neighbors are the same as before;
e.g. $-J_{01}\tilde{\sigma}^z_{12}\sigma_0^z$, and
the new effective transverse field acting to flip
the spin cluster is 
\begin{equation}
\tilde{h}_{12}\approx\frac{h_1h_2}{J_{12}}<J_{12}
\label{eq:A4}
\end{equation}
by second order perturbation theory. Since the two
spins in the cluster will tend to flip together,
their correlations will be strong
\begin{equation}
\langle \sigma^z_1\sigma^z_2 \rangle \approx1-O\left(
\frac{h_1}{J_{12}},\frac{h_2}{J_{12}}
\right).
\label{eq:A5}
\end{equation}

By decimating either a spin or a bond we have
decreased the number of degrees of freedom in the
system. The resulting effective Hamiltonian has
exactly the same form as before but we now have
spin clusters and effective bonds with lengths
$\{l_{Ci}\}$ and $\{l_{Bi}\}$, respectively, which
now vary.  This decimation procedure is iterated
with the largest of the couplings in the new
effective Hamiltonians
$\Omega\equiv\mbox{max}\{\tilde{h}_i,\tilde{J}_i\}$
gradually being reduced.

This renormalization group transformation has two
essential features.  First, the effective bonds  and
fields  remain {\it independent} (although, e.g.
the length of a bond and its strength {\it are}
correlated). Second, if the system is near
critical, the distributions of the basic {\it
logarithmic} couplings:
$$
\zeta_i\equiv\ln (\Omega/\tilde{J}_i)
$$
and
\begin{equation}
\beta_i \equiv\ln (\Omega/\tilde{h}_i)
\label{eq:A6}
\end{equation}
become {\it broader and broader}---even on this
logarithmic scale---due to the additive nature of
the recursion relations for $\zeta$ and $\beta$;
e.g. from Eq~(\ref{eq:A2}), 
\begin{equation}
\tilde{\zeta}_{13}=\zeta_{12}+\zeta_{23}.
\label{eq:A7}
\end{equation}
This implies that at low energy scales,
$\Omega \ll \Omega_I$, corresponding to 
\begin{equation}
\Gamma\equiv-\ln \Omega
\label{eq:A8}
\end{equation}
large, the perturbation expansion used in deriving
Eqs.~(\ref{eq:A2}-\ref{eq:A5}) becomes very good as
the neighboring couplings of a decimated coupling
(which is by prescription equal to $\Omega$) will
typically have magnitude $ \ll \Omega$. The occasional
bad case in which two neighboring couplings are
comparable can be shown (see F Section VI) to lead
only to corrections to scaling and to non-universal
amplitudes.  

The natural quantities to work with
are the joint distributions of the logarithmic
couplings and lengths $P(\zeta, l_C)$ and
$R(\beta, l_B)$ of the bonds and clusters,
respectively. The non-linear integro-differential
recursion relations for these (F2.13) and F(2.14)
are simply obtained. In terms of the Laplace
transform variables, 
\begin{equation}
P(\zeta, y)\equiv\int^\infty_0\  dle^{-yl}P(\zeta,
l)
\label{eq:A9}
\end{equation}
the recursion relations for each $y$  {\it
decouple} due to the simple additive properties of
lengths.  As shown in F Sec II, there is a special
four parameter family (for each $y$)  of exact
solutions of the RG recursion relations.  General
well behaved initial distributions converge very
rapidly to this special family and then at low
energy scales converge to a subfamily of {\it scaling
solutions} parametrized simply by $\delta$, defined in Eq.~(\ref{delta}), and
and a
basic length scale, $l_v$, defined in Eq.~(\ref{lv}).
%\begin{equation}
%l_v=\frac{2}{\mbox{var}_I(\ln
%J)+\mbox{var}_I(\ln h)}
%\label{eq:A10}
%\end{equation}
%where the subscripts on the variances denote the
%initial distributions.
Note that the particular
initial distributions we have used for our
numerical studies are in the special family at the
critical point but not away from it; this will
result only in corrections to scaling.
%For this
%choice of distributions the scale 
%\begin{equation}
%l_v=1,
%\label{eq:A11}
%\end{equation}
%as can be seen from Eq~(\ref{dist}).

\subsection{Obtaining End--to--End Correlations and
Gaps.}

\begin{figure}
\epsfxsize=\columnwidth\epsfbox{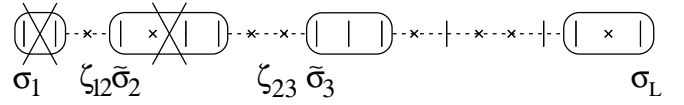}
\caption{Schematic of renormalization group decimation and its
use  for obtaining end-to-end correlations.  Clusters of spins (vertical
lines) are demarcated by closed curves, decimated spins are indicated
by small x's and decimated clusters by large X's. The dotted lines
denote the effective bonds coupling the spin clusters.  At a scale
$\Gamma_1$, the cluster containing the first spin $\sigma_1$ was
decimated with an effective bond strength
$e^{-\Gamma_1-\zeta_{12}}$ coupling it to the left-most remaining
cluster with spin $\tilde\sigma_2$. Later, at scale
$\Gamma_2>\Gamma_1$, the $\tilde\sigma_2$ cluster was decimated, and
the left-most remaining cluster is $\tilde\sigma_3$ as shown.  The
last spin, $\sigma_L$, is in a cluster that has not yet been decimated.
}
\label{dsf1}
\end{figure}

In order to obtain the properties of a finite
chain of length $L$ with free boundary conditions,
we must follow the correlations of the first and
last spins $\sigma^z_1$ and $\sigma^z_L$ with the
active---i.e. undecimated---spins that remain at
log-energy scale $\Gamma$. Since we will always be
interested in the $\sigma^z$ correlations, we will
drop the $z$ superscript for convenience. As
discussed above and in detail in F Sec VB, the
correlations of an end spin, say the
first, $\sigma_1$, can be obtained by first
observing that just after it is decimated, say at
a scale we denote
$\Gamma_1$, its correlations with the spins in the
leftmost remaining cluster, which has cluster spin
$\tilde{\sigma}_2(\Gamma_1)$, are given by
Eq~(\ref{eq:A3}) with the appropriate effective
bond at that scale
$\tilde{J}_{12}=e^{-\Gamma_1-\zeta_{12}}$.
and effective field  $\tilde{h}_1=e^{-\Gamma_1}$ on
the just-decimated cluster   which the $\sigma_1$
was in, i.e. 
\begin{equation}
\langle \sigma_1\tilde{\sigma}_2(\Gamma_1) \rangle =
e^{-\Lambda_F(\Gamma_1)}
\label{eq:A12}
\end{equation}
with 
\begin{equation}
\Lambda_F(\Gamma_1)\approx\zeta_{12}.
\label{eq:A13}
\end{equation}
Note that we have ignored here the corrections at
earlier stages of the renormalization from
Eq~(\ref{eq:A5}) as these will only give rise to
multiplicative factors of order unity,
corresponding to additive factors of order unity
in $\Lambda$, that arise from high energy scales
$\Omega\raisebox{-.6ex}{
$\stackrel{\textstyle<}{\sim}$ }\Omega_I$. These
are only corrections to scaling. At a later stage
of the RG, just after the leftmost cluster is
itself decimated at scale $\Gamma_2$, its
correlation with the new leftmost remaining
cluster $\tilde{\sigma}_3(\Gamma_2)$ is determined
by the effective bond connecting them which has
magnitude given by $\zeta_{23}$. Therefore from
Eq~(\ref{eq:A3})
\begin{equation}
\Lambda_F(\Gamma_2)\equiv-\ln
\langle \sigma_1\tilde{\sigma}_3(\Gamma_2) \rangle 
\approx \Lambda_F(\Gamma_1)+\zeta_{23}
\label{eq:A14}
\end{equation}
This process is illustrated schematically in
Fig.~\ref{dsf1}.

We can continue the process from both
ends until there is only one remaining spin
cluster with effective field
\begin{equation}
\tilde{h}_E=e^{-\Gamma_E}.
\label{eq:A14aa}
\end{equation}

The log-correlations of the first and last spins
with this last cluster are given by the $\Lambda$'s
at this scale, $\Lambda_F(\Gamma_E)$ and
$\Lambda_L(\Gamma_E)$ (for ``first'' and ``last'')
respectively.  We thus have, for the end-to-end
correlation function of interest,
\begin{equation}
\Lambda\equiv-\ln ( C_{1,L} ) =\Lambda_F(\Gamma_E)
+\Lambda_L(\Gamma_E). 
\label{eq:A14a}
\end{equation}
with additive errors typically of order unity.
The energy gap is simply given by the scale
$\Gamma_E$ at which the last degree of
freedom---this last cluster---is decimated:
\begin{equation}
G\equiv-\ln (\Delta E)\approx\Gamma_E.
\label{eq:A15}
\end{equation}

In order to obtain the distributions of these
quantities of interest, we must keep track of the
joint distribution at scale $\Gamma$ of the number
of remaining clusters, $N$, their $N$
log-effective fields $\{\beta_i\}$, the $N-1$
log-bond strengths $\{\zeta_i\}$ and the
contributions to the log-correlations of this
stage of the RG, $\Lambda_F(\Gamma)$ and
$\Lambda_L(\Gamma)$ from Eq~(\ref{eq:A14}), etc. A
natural first guess is that, as in the infinite
system, each of the couplings will be independent.
But, as shown explicitly in F Section IID, this is
{\it not} correct: correlations arise from the
constraint that the total length is $L$. For
example an anomalously small coupling, since it
tends to be that of a long cluster or bond, will
imply other shorter-than-typical segments with 
larger-than-typical couplings.  

Fortunately, a
modification of the naive guess of independence
turns out to be correct, but it involves explicitly
keeping track of the length of each
segment---including the first and last, $l_F$ and
$l_L$, that link
$\sigma_1$ and
$\sigma_L$ to the first and last remaining clusters
at scale
$\Gamma$. One can show that a distribution of the
form \cite{dprob}
\begin{eqnarray}
%\lefteqn{d\mbox{Prob} [N; \Lambda_F, l_F;
& & d\mbox{Prob} [N; \Lambda_F, l_F;
%\beta_1,l_{C1};\zeta_1, l_{B1} \dots }\label{eq:A16} \nonumber \\
\beta_1,l_{C1};\zeta_1, l_{B1} \dots \nonumber \\
& & \qquad \zeta_{N-1}, l_{BN-1}; \beta_N, l_{CN};\Lambda_L,
l_L\mid L,\Gamma]\nonumber\\
%\lefteqn{=a_N(L,\Gamma){\cal L}(\Lambda_F, l_F)
& & =  a_N(L,\Gamma){\cal L}(\Lambda_F, l_F)
d\Lambda_F dl_F R(\beta_1,l_{C1}) d\beta_1
%dl_{C1}}\nonumber\\
dl_{C1}\nonumber\\
& & \qquad P(\zeta_1,l_{B1})d\zeta_1  dl_{B1}  \times\dots
\times {\cal L}(\Lambda_L, l_L) d\Lambda_L dl_L \nonumber \\
& & \qquad 
\delta(l_F+l_{C1}+l_{B1}+\dots+l_{CN}+l_L-L)
\label{eq:A16} 
\end{eqnarray}
for $N > 1$ has its form preserved under
renormalization if 
\begin{equation}
a_N(L,\Gamma)=a(L,\Gamma)
\label{eq:A17}
\end{equation}
is independent of $N$ with
\begin{equation}
\frac{1}{a} \frac{da}{d\Gamma}=\int^\infty_0 \ dl [
P(\zeta=0,l) + R(\beta=0,l) ].
\label{eq:A18}
\end{equation}
Note that the functions ${\cal L}$, $P$ and $R$ and
all other functions that will occur later are
implicitly functions of  $\Gamma$ unless otherwise
stated.  

Initially, $N=L$ is the only term with non-zero
probability, and all the $\{l_{Bi}\}$ and
$\{l_{Ci}\}$ are equal to $\frac{1}{2}$, (a
convention chosen for convenience).
Therefore the needed solution has exactly the form
of Eq~(\ref{eq:A16}).  In terms of functions 
${\cal L}$, $P$, and $R$ of the Laplace transform
variable, $y$, the constraint can simply be enforced
via the inverse Laplace transform
\begin{equation}
\raisebox{-1.8ex}{
$\stackrel{\textstyle LT^{-1}}{y\to
L}$}\equiv\frac{1}{2\pi i}\int dy\ e^{yL}
\label{eq:A19}
\end{equation}
at the end of a computation. Normalization requires
that initially
\begin{equation}
a(L,\Gamma_I)=\frac{1}{n_{\Gamma_I}}
\label{eq:A20}
\end{equation}
(which is unity in our case) independent of $L$.
>From Eqs.~(F2.16) and (\ref{eq:A18}) we then have
\begin{equation}
a(\Gamma)=\frac{1}{n_\Gamma}
\label{eq:A21}
\end{equation}
for all $\Gamma$ where $n_\Gamma$ is the number
density per unit length of the clusters remaining 
at scale
$\Gamma$ in an infinite system.

The information we are interested in can be
obtained from the function 
\begin{eqnarray}
\label{eq:A22}
Z(\Lambda, L, \Gamma) d\Lambda\equiv d\mbox{Prob }
[\mbox{no cluster remaining}\\
\mbox{ and }-\ln C_{IL}=\Lambda\mid L,\Gamma].\nonumber
\end{eqnarray}
When the last cluster is decimated at $\Gamma_E$,
this will yield a contribution to 
$Z[\Lambda=\Lambda_F(\Gamma_E)+\Lambda_L(\Gamma_E)
,L,\Gamma_E]$. In the Laplace transform variable $y$
for the lengths, using Eq~(\ref{eq:A16}) we have 
\begin{eqnarray}
\lefteqn{J(\Lambda,y,\Gamma)\equiv
\raisebox{-1.8ex}{
$\stackrel{\textstyle LT^{-1}}{L\to
y}$}\  \frac{\partial Z(\Lambda,L,\Gamma)}{\partial
\Gamma}}\label{eq:A23}\\
&= &  a(\Gamma)\int d\Lambda_F \int d\Lambda_L
\delta(\Lambda-\Lambda_F-\Lambda_L)\nonumber\\
& &R(\beta=0,y){\cal L}(\Lambda_F,y){\cal
L}(\Lambda_L,y)
\nonumber
\end{eqnarray}
since the probability of decimation of the last
cluster with length $l_C$ as $\Gamma$ is increased
by $d\Gamma$ is $R(\beta=0,l_C)d\Gamma$. The
function 
\begin{equation}
J(\Lambda, L, \Gamma=G)\equiv\raisebox{-2ex}{
$\stackrel{\textstyle LT^{-1}}{y\to
L}$}\ J(\Lambda,y,\Gamma=G)
\label{eq:A24}
\end{equation}
is then the joint probability distribution of
$\Lambda=-\ln C_{1,L}$ and $G=-\ln\Delta E.$

The only unknown function in this process is
${\cal L}(\Lambda,l;\Gamma)$. It is just the joint
distribution of the log-correlations of the first
spin in a semi-infinite chain with the leftmost
undecimated spin cluster at scale $\Gamma$ and
the length of the bond connecting these.
The Laplace transform of ${\cal L}$ in $L\to y$ and
$\Lambda\to z$ is simple to analyze given the
additive form of the recursion relation
Eq~(\ref{eq:A14}). This function ${\cal L}(z,y=0)$
(i.e. ignoring the $l$ variable) is exactly the
function ${\cal L}(z)$ in F Section VB.  Following
the analysis presented there we have 
\begin{equation}
\frac{\partial{\cal L}(z,y)}{\partial\Gamma} =
{\cal L}(z,y)[P(z,y) R(\beta=0,y)-R(\beta=0, y=0)]
\label{eq:A27}
\end{equation}
which can be integrated.  Using the
special family of exact solutions for $P$ and $R$,
we obtain
\begin{equation}
{\cal
L}(z,y)=\frac{u(0)[z+u_I(y)]}{u_I(0)[z+u(y)]} 
\label{eq:A28}
\end{equation}
with
\begin{displaymath}
u(y)\equiv u(y,\Gamma),
%\label{eq:A29}
\end{displaymath}
and
\begin{equation}
u_I(y)\equiv u(y,\Gamma_I)
\label{eq:A29}
\end{equation}
%\begin{eqnarray}
%& &u(y)\equiv u(y,\Gamma), \label{eq:A29} \\
%\mbox{and}\nonumber\\
%& &u_I(y)\equiv u(y,\Gamma_I)\nonumber
%\end{eqnarray}
given generally by Eqs.~(F2.51) and (F2.52) which we
will not need in detail here. From Eqs.~(F2.46) and
(F2.53) we have
\begin{equation}
R(\beta=0,y)=T(y), 
\label{eq:A30}
\end{equation}
defined below,
and hence from Eq~(\ref{eq:A23}),
\begin{equation}
J(z,y,\Gamma)=\frac{1}{n_\Gamma}T(y) \left[
\frac{z+u_I(y)}{z+u (y)}
\right]^2\frac{u(0)^2}{u_I(0)^2}
\label{eq:A31}
\end{equation}
In the scaling limit of large $\Gamma$, small $y$, 
\begin{eqnarray}
& &u(y)=\Delta(y)\coth[\Gamma\Delta(y)]-\delta\label{eq:A32}\\
& &T(y)=\frac{\Delta (y) e^{\delta \Gamma}}
{\sinh[\Gamma\Delta(y)]}\nonumber
\end{eqnarray}
where
\begin{equation}
\Delta(y)=\sqrt{y+\delta^2}
\label{eq:A33}
\end{equation}
with lengths measured in units of $l_v$. Since the
scales involved initially are small, for small $y$
we have
\begin{equation}
u_I(y)\approx u_I(0).
\label{eq:A34}
\end{equation}
In the scaling limit,
\begin{equation}
n_\Gamma\approx
\frac{\delta^2}{\sinh^2(\Gamma\delta)}\ .
\label{eq:A35}
\end{equation}

Since in the scaling limit we expect $\Lambda$ to
typically be large,  its distribution will be
determined by the small $z$ behavior, so we can
also write $z+u_I(y)\approx u_I(0)$; [We will  
consider later what has been left out by this
approximation.] We now see that the
$u_I(0)$'s cancel in the scaling limit yielding
\begin{equation}
J(z,y,\Gamma)=\frac{\Delta e^{-\delta\Gamma} }
{\left[
 z-\delta+\Delta\coth(\Gamma\Delta) \right]^2
\sinh (\Gamma\Delta)}
\label{eq:A36}
\end{equation}
with $\Delta(y)=\sqrt{y+\delta^2}$ and 
\begin{equation}
d\mbox{Prob}(\Lambda, G\mid L)=d\Lambda dG 
\raisebox{-1.8ex}{
$\stackrel{\textstyle LT^{-1}}{y\to
L}$} \ \raisebox{-1.8ex}{
$\stackrel{\textstyle LT^{-1}}{z\to
\Lambda}$} J(z,y,\Gamma=G)
\label{eq:A37}
\end{equation}
our principal result.

\subsection{Distribution of Correlations}

The distribution of the log of the end-to-end correlations,
$\Lambda \equiv -\ln C_{1,L}$,
can be obtained by integrating Eq~(\ref{eq:A37})
over $G$. The inverse Laplace transform in $z$ can
be done trivially yielding
\begin{equation}
J(\Lambda,y,\Gamma)=\Lambda e^{\delta\Lambda
-\Lambda\Delta\coth(\Gamma\Delta)} \frac {\Delta}
{\sinh(\Gamma\Delta)}
e^{-\delta\Gamma}\Theta(\Lambda).
\label{eq:A38}
\end{equation}
{\it At criticality}, $\delta=0$, we can write
\begin{equation}
w=\frac{1}{\sinh(\Gamma\Delta)}
\label{eq:A39}
\end{equation}
and convert the $\Gamma$ integral to a $w$
integral. The inverse Laplace transform in $y$ can
then be done straightforwardly yielding a simple
integral over $w$ and thereby 
\begin{eqnarray}
d\mbox{Prob}(\Lambda\mid L,
\delta=0)&=&d\Lambda
\frac{\Lambda}{2L} e^{-\Lambda^2/4L}\label{eq:A40}\nonumber\\
&=& d\lambda 
\raisebox{.5ex} {${\scriptstyle\frac{1}{2}}$}
\lambda e^{-\lambda^2/4}
\end{eqnarray}
which is of scaling form in
\begin{equation}
\lambda\equiv\frac{\Lambda}{\sqrt{L}}.
\label{eq:A41}
\end{equation}

{\it Off-critical}, we have not managed to compute 
the distribution of $\Lambda$ in closed form. But
progress can be made on the {\it moments} by
expanding $J$ in powers of $z$. 

In the {\it disordered
phase}, for small fixed $\delta>0$ and small $y$, we
can expand $\Delta\approx
y\delta+\frac{1}{2}\frac{y}{\delta}+\cdots$. For
large $L$, one then obtains 
\begin{equation}
\overline{\Lambda^p}\approx(2\delta L)^p
\label{eq:A42}
\end{equation}
so that the distribution is strongly peaked around
$2\delta L$. For $p=1$, Eq~(\ref{eq:A42}) is the
first term in an expansion in powers of $\xi/L$
with the correlation length
\begin{equation}
\xi\approx\frac{l_v}{\delta^2}.
\label{eq:A43}
\end{equation}
To next order we obtain
\begin{equation}
\overline{\Lambda}\approx 2\frac{\delta L}{l_v}
\left\{ 1+\frac{\xi}{4L} \left[  \ln \left(
\frac{4L}{\xi} \right) + \gamma
\right] +O\left(\frac{\xi^2}{L^2} \right)
 \right\}
\label{eq:A44}
\end{equation}
with $\gamma\approx 0.577$, Euler's constant. The
variance of $\Lambda$ is
\begin{equation}
\mbox{var}(\Lambda)=\overline{(\Lambda-
\overline{\Lambda})^2}\approx2 \frac{L}{l_v}+o
(L)
\label{eq:A45}
\end{equation}
with the correction probably a power of $\ln L$.
In Eqs.~(\ref{eq:A43})-(\ref{eq:A45}) we have
reinserted factors of the basic length $l_v$.

Note, from Eq.~(\ref{eq:A44}), that $\overline{\Lambda}$ is of order one for
$L= \tilde{\xi}$, where
\begin{equation}
\tilde{\xi} = {l_v \over 2 \delta}
\label{xi_typ}
\end{equation}
is the {\em typical} correlation length. However, the {\em true} correlation
length, $\xi$ in Eq.~(\ref{eq:A43}), is the scale at which
%that the correlation length $\xi$ is the scale
%at which 
\begin{equation}
\overline{\Lambda}(L=\xi)\sim\sqrt{\mbox{var}\,
[\Lambda (L=\xi)]},
\label{ea:A48}
\end{equation}
which is much larger.
%than the scale at which
%$\overline{\Lambda}$ is of order one.
Thus the
typical correlations are very small at distances
of order the correlation length.

In the {\it ordered phase}, $\delta<0$, for long
chains the end spins will almost have a
spontaneous magnetization. We thus expect that
$C_{1,L}$ will be approximately the product of two
independent  end-point  magnetizations of an
infinite system.  From F section VB, we thereby obtain 
\begin{equation}
d\mbox{Prob}[\Lambda\mid L]\approx
4\delta^2\Lambda
e^{-2\mid\delta\mid\Lambda}d\Lambda.
\label{eq:A49}
\end{equation}
This can also be computed from $J$ using the fact
that for $\delta<0$, ${\cal L}(z,
y=0,\Gamma\to\infty)$ tends to a limit as seen in
F section VB.

\subsection{Energy gaps}

We now turn to the distribution of energy gaps,
more precisely of $G \equiv -\ln\Delta E$. This can be
obtained from Eq~(\ref{eq:A36}) by setting $z=0$,
$\Gamma=G$, and inverse Laplace transforming over
$y$. For general $\delta$, we have not been able
to perform this inverse transform analytically.
But at the {\it critical point} the behavior again
simplifies. By changing to the variable
$\Delta=\sqrt{y}$, the inverse transform can be done
by deforming the contour to circle a series of
double poles, yielding, in terms of the scaled
variable
\begin{equation}
g\equiv\frac{G}{\sqrt{L}}
\label{eq:A50}
\end{equation}
\begin{eqnarray}
d\mbox{Prob}[g&\mid& L, \delta=0] \equiv dg
D(g,\delta=0) \nonumber \\
&=& dg \frac{2\pi}{g^3} \sum^\infty_{n=-\infty} 
e^{-(n+\frac{1}{2})^2\pi^2/g^2}(-1)^n(n+\frac{1}{2})
\nonumber\\  &=& dg \frac{2}{\sqrt{\pi}}
\sum^\infty_{m=-\infty}
e^{-g^2(m+\frac{1}{2})^2
}(-1)^m(m+\frac{1}{2})
\label{de_scale}
\end{eqnarray}
the second expression being the Poisson resummed
form of the first. We see that for large $g$, 
\begin{equation}
D(g,\delta=0)\approx\frac{2}{\sqrt{\pi}}e^{-g^2/4};
\label{eq:A52}
\end{equation}
while for small $g$,
\begin{equation}
D(g,\delta=0)\approx\frac{2\pi}{g^3}e^{-\pi^2/(4g^2)}.
\label{eq:A53}
\end{equation}
Thus the probability of both very large and very
small gaps is exponentially small.  The former implies that the average gap will 
{\it not} decay as a power of $L$.  Nevertheless, from Eq(\ref{de_scale}) one 
can see that the  integration over $g$ needed to obtain the average gap is
dominated by the maximum of the  integrand yielding 
\begin{equation}
 \overline{ \Delta E}   \sim L^{\frac{1}{6}} \exp
\left(-\frac{3}{2}\left(\frac{\pi ^2  L}{2}\right)^{\frac{1}{3}}\right)
\label{av-gap}
\end{equation}
where we have dropped a (non-universal) prefactor.  Thus at the 
critical point the average gap decays more slowly with $L$---indeed with a 
different form of $L$ dependence---than does a typical gap which has the form
$\sim
\exp( \rm{const.} L^{1/2})$.

In the {\it disordered phase}, the behavior for
$L \gg \xi$ can be obtained by expanding
$\coth(\Gamma\Delta)$ in $y$ yielding 
\begin{equation}
J(z=0,y,\Gamma=G)\approx\frac {2\delta
e^{-2\delta G}} {\left(\frac{y}{2\delta}+2\delta
e^{-2G\delta}\right)^2}
\label{eq:A54}
\end{equation}
valid for
\begin{equation}
1 \ll \delta G \ll L\delta^2=\frac{L}{\xi}.
\label{eq:A55}
\end{equation}
Noting that in this limit
\begin{equation}
n_G\approx 4\delta^2 e^{-2\delta G},
\label{eq:A56}
\end{equation}
the inverse Laplace transform simply yields
\begin{eqnarray}
d\mbox{Prob}[G\mid L,\delta>0]&=&J(z=0, L, \Gamma=G)
dG\label{eq:A57}\\  
&\approx& 2\delta n_G Le^{-Ln_G}dG.\nonumber
\end{eqnarray}
This result has a simple interpretation
\cite{yr}.

For energy scales outside of
the critical region, i.e.
$\Gamma \gg \frac{1}{\delta}$, the remaining clusters
are typically of length small compared to their
separations, the latter being typically of order
$1/n_G$. 
These can be thought of\cite{yr} as independent localized
excitations with density $n_G$. Thus the probability
of one of those in a sample having an anomalously
small
$G$ is proportional to the sample size $L$. For
fixed
$L$ and 
\begin{equation}
\delta
L \gg G \gg \frac{1}{\delta}\ln\left(\frac{L}{\xi}\right),
\label{eq:A59}
\end{equation}
Eq~(\ref{eq:A57}) is seen to have exactly this
form.
Since, from Eq.~(\ref{eq:A56}), $n_G \sim (\Delta E)^{2\delta}$, 
the distribution 
%with an implied distribution
of the energies
of a single cluster has the form
\begin{equation}
\mbox{Prob}[(\Delta E)_{cluster}<\varepsilon]\sim
\varepsilon^{1/z}
\label{eq:A60}
\end{equation}
with
\begin{equation}
z=\frac{1}{2\delta}.
\label{eq:A61}
\end{equation}
This interpretation \cite{yr}
assumes
that clusters are separated by of order $\xi$ as
they indeed are at the crossover scale  away from
criticality, $\Gamma\sim1/\delta$. In this regime,
length $L$ corresponds to
$(\mbox{energy})^{1/z}$ so that $z$ plays the
role of a continuously variable dynamic exponent as
discussed in F section IVA and Ref.~\onlinecite{yr}.  Indeed,
Eq~(\ref{eq:A57})
has exactly  the form of a
scaling function of
\begin{equation}
\hat{\varepsilon}\sim\varepsilon\left( \frac{L}
{\xi} \right)^z
\label{eq:A62}
\end{equation}
valid for $\frac{L}{\xi}\to \infty$ and
$\varepsilon\to 0$ with $\hat{\varepsilon}\to$ any
fixed value.
Note that for extremely small gaps, $G \gg \delta L$,
(for which the above results do not hold)  the
distribution depends on extremely rare clusters and
can become less universal.

In the {\it ordered phase}, $\delta<0$, the
distribution of gaps for
$L \gg \xi=\frac{1}{\delta^2}$ can be found simply. If
we guess that $G$ will typically be of order $L$,
then the integral over $y$ can be done by steepest
descents with the saddle point at
$\Delta(y)\approx\frac{\Gamma}{2L}$. The
distribution of $G$ is then found to be
asymptotically Gaussian for
$L \gg \frac{1}{\delta^2}$:
\begin{equation}
d\mbox{Prob}(G\mid L, \delta<0)\approx \frac{dG}
{\sqrt{4\pi L}}e^{-(G-2\mid\delta\mid L)^2/4L}.
\label{eq:A63}
\end{equation}
We thus have 
\begin{equation}
\overline{G}=2\mid\!\delta\!\mid L/l_v
\label{eq:A64}
\end{equation}
and
$$\mbox{var}\, G\approx 2L/l_v
$$
where we have reinserted factors of $l_v$.

\subsection{Relation between correlations, gaps
and couplings}

We have seen that in the ordered phase the
logarithm of the gap, $-G$, and in the disordered
phase the logarithm of the end-to-end
correlations, $-\Lambda$, both have 
approximately Gaussian statistics in long samples;
i.e. $L \gg \xi$. Indeed, for $\delta<0$ in this limit
\begin{equation}
\frac{\overline{G}}{L}\approx { 2\mid\!\delta\!\mid \over  l_v}
=\overline{\ln J}-\overline{\ln h}
\label{eq:A65}
\end{equation}
and
$$\frac{\mbox{var}(G)}{L}\approx\frac{2}{l_v} =
\mbox{var}(\ln J)+\mbox{var}(\ln h),
$$
while similarly for $\delta>0$,
\begin{eqnarray}
\frac{\overline{\Lambda}}{L}&\approx& { 2\delta \over l_v}
=\overline{\ln h}-\overline{\ln J} \label{eq:A66}  \\
\frac{\mbox{var}(\Lambda)}{L}&\approx&\frac{2}{l_v}
=\mbox{var}(\ln J)+\mbox{var}(\ln h).\nonumber
\end{eqnarray}
Where does this simple behavior come from?
Examination of the basic recursion relations
Eqs.~(\ref{eq:A2}) and (\ref{eq:A4}) imply that at
any scale, the effective field on any cluster is
simply
\begin{equation}
\tilde{h}\approx e^{-\Sigma_C}
\label{eq:A67}
\end{equation}
and the strength of any effective bond,
\begin{equation}
\tilde{J}\approx e^{-\Sigma_B}
\label{eq:A68}
\end{equation}
with
\begin{equation}
\Sigma_{\rm section}\equiv\sum_{i \epsilon\ \rm{section}}(\ln h_i-
\ln J_i)
\label{eq:A69}
\end{equation}
the sum 
running over all the original bonds or
spins in the cluster ($\Sigma_C$) or effective bond
($\Sigma_B$).  The last cluster to be decimated in a
chain with free boundary conditions thus
determines  {\it both} the gap and the end-to-end
correlations. If $\Sigma_{EC}$ is the sum of
Eq~(\ref{eq:A69}) over the last to be decimated
cluster and $\Sigma_T$ is the sum over the {\em whole}
chain, i.e.
\begin{equation}
\Sigma_T = \sum_{i=1}^L  \ln h_i -
\sum_{i=1}^{L-1} \ln J_i,
\label{Sigma_T}
\end{equation}
then
\begin{equation}
G\approx-\Sigma_{EC}
\label{eq:A70}
\end{equation}
since the gap is the effective field on the last
cluster. Likewise, the log-correlations are given
by
\begin{equation}
\Lambda\approx\Sigma_T-\Sigma_{EC}
\label{eq:A71}
\end{equation}
since, from Eq~(\ref{eq:A14}),
$\Lambda_{F}(\Gamma_{E})$ will be the sum of $\ln
h_i-\ln J_i$ over the section to the left of the
last cluster and likewise
$\Lambda_{F}(\Gamma_{E})$ to the right.
Eqs.~(\ref{eq:A70}) and (\ref{eq:A71}) immediately give the simple result that
\begin{equation}
\Lambda - G \approx \Sigma_T .
\label{eq:L-G}
\end{equation}
Note that
the RG procedure guarantees the physical requirements that $\Lambda\geq 0$
and 
\begin{equation}
G\geq-\ln\Omega_I.
\label{eq:A72}
\end{equation}
The specific set of bonds and spins which make up 
the last cluster for a given sample are thus very
special.

In the disordered phase  for
$\Gamma \gg \frac{1}{\delta}$ and $L \gg \xi$, the
remaining clusters are typically small [of
length $\xi$ times factors like $\ln(L/\xi)$]. 
Thus as $L/\xi\to\infty$,
\begin{equation}
\overline{\Lambda}\approx\overline{\Sigma_T}+o(L)
\label{eq:A73}
\end{equation}
and
\begin{equation}
\mbox{var}\,\Lambda_T\approx\mbox{var}\,\Sigma_T+o(L).
\label{eq:A74}
\end{equation}
Similarly, in the ordered phase most of the system
will be in one cluster; this is the
reason for spontaneous magnetization in an
infinite system. Thus $\Sigma_{EC}\approx\Sigma_T$
so that
\begin{eqnarray}
\overline{G}\approx\overline{\Sigma_T}+o(L)\\
\mbox{var}\,G\approx\mbox{var}\,\Sigma_T+o(L).\nonumber
\end{eqnarray}
The results Eqs.~(\ref{eq:A65})-(\ref{eq:A66})
immediately follow.

A check on this picture can be obtained by
computing the distribution of $\Lambda-G$. From 
Eq.~(\ref{eq:A36}) 
this is found to be purely
Gaussian with the mean and variance identical
(within the RG approximation) to that of
$\Sigma_T$, as also follows directly from (\ref{eq:L-G}). 
>From Eqs.~(\ref{Sigma_T}) and (\ref{eq:L-G}) the
central limit theorem tells us that the distribution of $\Lambda-G$ is
Gaussian with mean and variance given by
\begin{eqnarray}
\overline{\Lambda - G} & = & L(\Delta_h - \Delta_J) =
{2 \delta \over l_v} L  \nonumber \\
\mbox{var}(\Lambda - G) & = & L(\mbox{var}\, h + \mbox{var}\, J) =
{2 \over l_v} L  ,
\label{L-G}
\end{eqnarray}
where we used Eqs.~(\ref{delta}) and (\ref{lv})

We conclude this section with a note on connection
to earlier work. Shankar and Murthy\cite{sm} studied the
correlations of quantities, such as
$\sigma^z_i\sigma^z_{i+1}$ or
$\sigma^x_i$ which are local in fermion variables.
Their results imply that, e.g., 
\begin{equation}
\frac{-\ln \langle \sigma^x_i\sigma^x_{i+r} \rangle }{\mid
r\mid}\to\mid \overline{\ln h}-\overline{\ln J}\mid
\label{eq:A75}
\end{equation}
with probability one for separation $\mid r
\mid\to\infty$. Thus the asymptotic decay of
``typical'' correlations in the disordered phase
of both $\sigma^z$ and $\sigma^x$, as well as the
local energy density etc., is exponential with the
{\it same} ``typical correlation length'', $\tilde{\xi}$ given by
Eq.~(\ref{xi_typ}). 
This prediction for the ``typical correlation length'' was confirmed
numerically in Ref.~\onlinecite{yr}.
%\begin{equation}
%\tilde{\xi}=\frac{1}{\overline{\ln h}-\overline{\ln
%J}}=\frac{l_v}{2\delta}.
%\label{eq:A76}
%\end{equation}

\subsection{Average end-to-end correlations}

The form of the average end-to-end correlation 
function, $\overline{C}_{1,L}$, can be obtained
much more simply than the distribution. Crudely, the
average is dominated by rare samples in which
neither of the two end spins is decimated until
the final scale $\Gamma_E$ when the last
cluster, which in this case contains both
$\sigma_1$ and $\sigma_L$, is decimated.
>From the analysis we have carried out above,
this corresponds simply to the $\delta(\Lambda)$
part of $J(\Lambda, L, \Gamma_E)$. Thus up to
non-universal factors arising from the high
energy small scale physics, we have
\begin{eqnarray}
\overline{C}_{1,L}&\sim& \raisebox{-1.8ex}{
$\stackrel{\textstyle LT^{-1}}{L\to
y}$} \int^\infty_{\Gamma I} d\Gamma 
J(z=0,y,\Gamma) \label{eq:A77} \\
& =  & \raisebox{-1.8ex}{
$\stackrel{\textstyle LT^{-1}}{L\to
y}$}\int^\infty_{\Gamma I} d\Gamma 
\frac{\Delta e^{-\delta\Gamma}}
{\sinh(\Gamma\Delta)}\frac{1}{u^2_I(0)}.\nonumber
\end{eqnarray}
At {\it criticality}, the integral over
$\Gamma$ is logarithmically divergent for small
$\Gamma$ yielding a $\ln (1/y)$ singularity for
small $y$ corresponding to 
\begin{equation}
\overline{C}_{1,L}\approx\frac{K}{L}
\label{eq:A78}
\end{equation}
for large $L$, 
where $K$ is a non-universal constant. The sources of this non-universality are
discussed in the next section.

\section{Numerical Results}
The numerical technique used here has been discussed previously\cite{yr,y,snm}.
%for periodic boundary conditions\cite{yr} and also
%for the simpler case of free boundary
%conditions\cite{y} which we consider here.
It involves mapping the
spin operators to fermions and diagonalizing the resulting {\em free} fermion
problem. We refer to earlier work\cite{yr,y} for details. We are able to study
sizes up to $L=256$ and obtain results for about 50,000 samples. This large
number is necessary in order to map out the distribution with some precision. 

\subsection{End-to-end Correlations}

\begin{figure}
\epsfxsize=\columnwidth\epsfbox{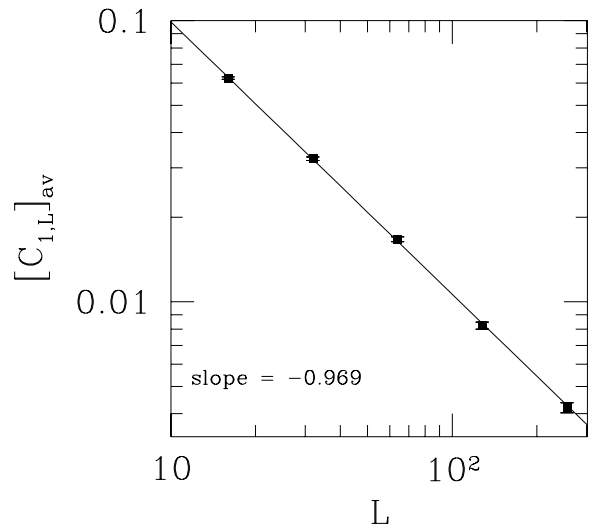}
\caption{
A log-log plot of 
the average of the end-to-end correlation function at the critical point.  The
solid line is the best fit and has a slope of 0.969, close to the prediction of
$-1$, see Eq.~(\protect\ref{eq:A78}).
}
\label{cfav}
\end{figure}

Our data for the average end-to-end correlation function is
shown in Fig.~\ref{cfav}.
The slope is about $-0.97$, close to the
prediction of $-1$ in Eq.~(\ref{eq:A78}),
and the amplitude, $K$, is close to unity.

\begin{figure}
\epsfxsize=\columnwidth\epsfbox{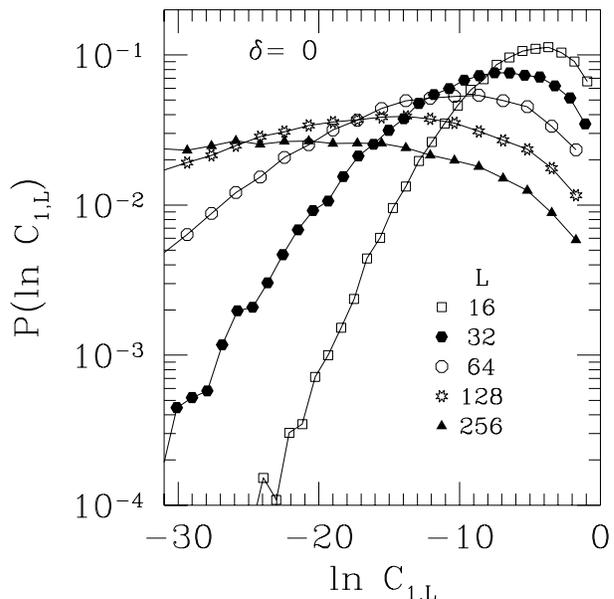}
\caption{
The distribution of the end-to-end correlation function at
the critical point for different sizes.
}
\label{cf_dist}
\end{figure}

\begin{figure}
\epsfxsize=\columnwidth\epsfbox{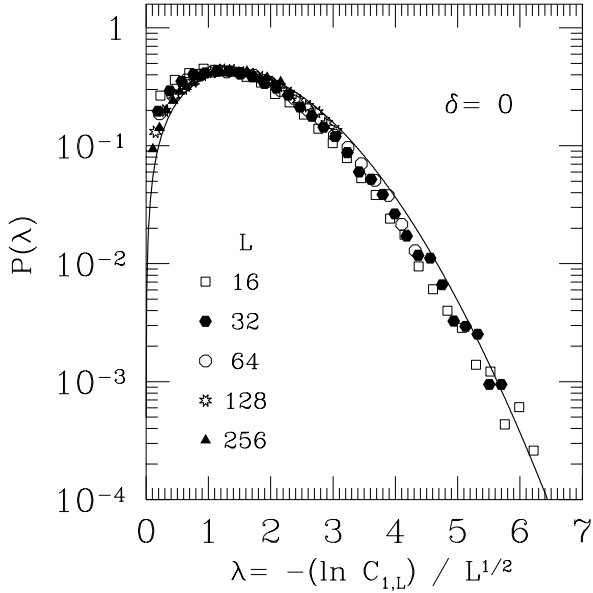}
\caption{
A scaling plot of the distribution of the end-to-end correlation function at
the critical point. The solid line is the analytical result in
Eq.~(\protect\ref{eq:A40}), $P(\lambda) = (\lambda/2) \exp( -\lambda^2/4)$.
}
\label{cf_dist_scale}
\end{figure}

The distribution of values of the end-to-end correlation 
function at the critical point is shown in
Fig.~\ref{cf_dist}. Clearly the distribution is extremely broad and becomes
broader with increasing size, even on a logarithmic scale.
The scaling plot in Fig.~\ref{cf_dist_scale} shows that the data fits
quite well the
analytical result of
Eq.~(\ref{eq:A40}).
%and (\ref{eq:A41}).
Note that there are no adjustable
parameters in this comparison. 

\begin{figure}
\epsfxsize=\columnwidth\epsfbox{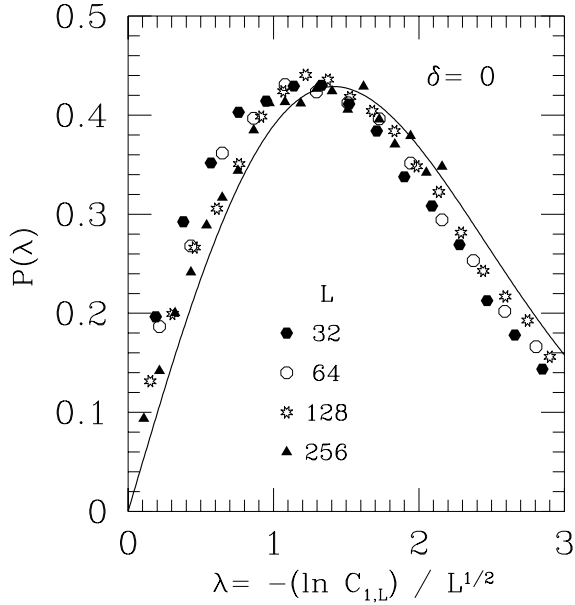}
\caption{
A blowup of the region of Fig.~\protect\ref{cf_dist_scale}
near the origin with a linear vertical scale.
}
\label{cf_dist_blowup}
\end{figure}

Fig.~\ref{cf_dist_blowup} shows the region near $\lambda=0$ with a linear 
vertical scale. Whereas the data for smaller sizes appear to have a finite
intercept at
$\lambda=0$ this decreases for larger sizes consistent with the 
analytic prediction, Eq.~(\ref{eq:A40}),
of a linear variation for small $\lambda$ in the scaling limit.

The average value in Eq.~(\ref{eq:A78})
arises from rare samples with exceptionally large values of the
correlation function  which correspond  to the small $\lambda$ region of
Figs.~\ref{cf_dist_scale} and \ref{cf_dist_blowup}. We shall see that 
the amplitude, $K$, of the $1/L$ behavior in Eq.~(\ref{eq:A78}) involves both
a contribution from the
 scaling function and a non-universal part which is not contained in the
scaling function. Noting that $C_{1,L}=\exp(-\lambda L^{1/2})$, 
the contribution to the
average from the scaling function is
\begin{equation}
[C_{1,L}]_{\rm av}^{\mbox{\scriptsize scaling part}}
= {1\over 2} \int_0^\infty e^{-\lambda 
L^{1/2}} \lambda  
e^{-\lambda ^2/4} \, d\lambda    .
\end{equation}
The dominant contribution is for $\lambda $ small, so we can replace
$e^{-\lambda ^2/4}$ by unity, which immediately gives
\begin{equation}
[C_{1,L}]_{\rm av}^{\mbox{\scriptsize scaling part}} = {1 \over 2L},
\label{scaling}
\end{equation}
corresponding to an amplitude, $K$, in Eq.~(\ref{eq:A78}) of 1/2, in
disagreement with the data in Fig.~\ref{cfav} which gives a value of about
unity.  In fact, the amplitude must clearly be non-universal, because one could
redefine $\sigma^z$ to be a multiple of itself which, from the scaling form in
the logarithmic variable $\lambda$, would change this amplitude but otherwise 
not
change the physics. Clearly then, one source of the
non-universal amplitude comes from the fact that 
$\ln C_{1,L}$ in the scaling function could be replaced
by $\ln(C_{1,L} / C_0)$, where $C_0$ is a
constant. The scaling regime involves the limit $\ln C_{1,L} \to -\infty$ and
so the factor of $C_0$ is a {\em correction}
to scaling (actually simply the leading correction to scaling discussed in F).
Note that including the factor of $C_0$ shifts the scaling curves in
Figs.~\ref{cf_dist_scale} and
\ref{cf_dist_blowup}
horizontally by an amount of order $1/L^{1/2}$, thus giving a finite intercept
at $\lambda =0$, which is seen in Fig.~\ref{cf_dist_blowup}.
Although the factor of $C_0$ is not part
of the scaling function, it does affect the average correlation function
since, repeating the calculation which led to Eq.~(\ref{scaling})
with $C_0$ included, obviously gives
\begin{equation}
[C_{1,L}]_{\rm av}^{\mbox{\scriptsize including amplitude}} = {C_0 \over 2L} .
\label{incl_s0}
\end{equation}

\begin{figure}
\epsfxsize=\columnwidth\epsfbox{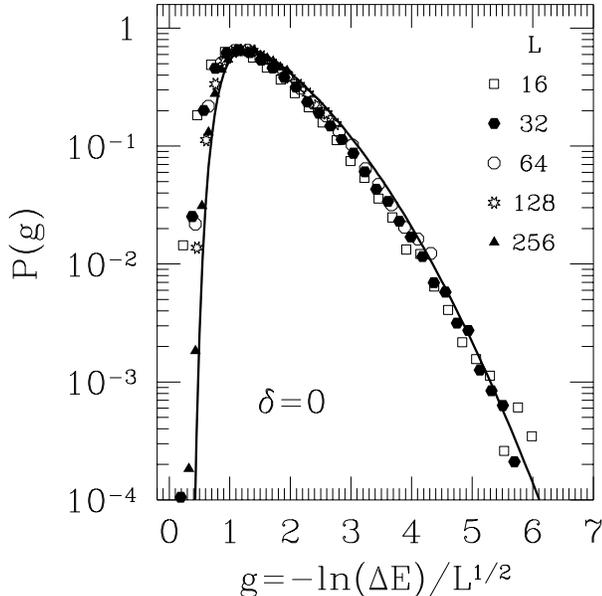}
\caption{
A scaling plot of the distribution of energy gaps at the critical point. The
solid line, which fits well, is a plot of Eq.~(\protect\ref{de_scale}) of the 
text.
}
\label{delta_E_scale}
\end{figure}

The factor of $C_0$
is not the only non-scaling contribution to the amplitude. 
In particular, we can go back and look at the non-scaling parts of the
distribution that arise from $J(z,y,\Gamma)$ in Eq.~(\ref{eq:A31}). The
scaling limit involved neglecting $z$ with respect to $u_I(y)$ in the
numerator of the factor in parenthesis in Eq.~(\ref{eq:A31}). But for large
$z$, which determines the behavior for small $\Lambda$, i.e. large
$C_{1,L}$, we should take the quantity in parentheses equal to unity. As
discussed earlier this corresponds to situations in which neither the spin
cluster containing
$\sigma_1$ nor that containing $\sigma_L$ is ever decimated: i.e. both end
spins remain active until the scale $\Gamma_E$ at which the last cluster,
which in this case contains both $\sigma_1$ and $\sigma_L$, is decimated.
The probability of this occurring is of order $1/L$ arising from the $u_I(0)$'s
in Eq.~(\ref{eq:A78}) but with a non-universal coefficient.

Thus we see that the mean correlations have various contributions of order
$1/L$ each with coefficients which depend on the high energy small scale
details that are {\it not} treated correctly in our RG analysis.

\subsection{Gaps}

In addition to calculating the end-to-end correlation function, 
we have also determined the energy gap at the critical point and show a
scaling plot of the results in Fig.~\ref{delta_E_scale}. The agreement with the
prediction in Eq.~(\ref{de_scale}) is good.
\begin{figure}
\epsfxsize=\columnwidth\epsfbox{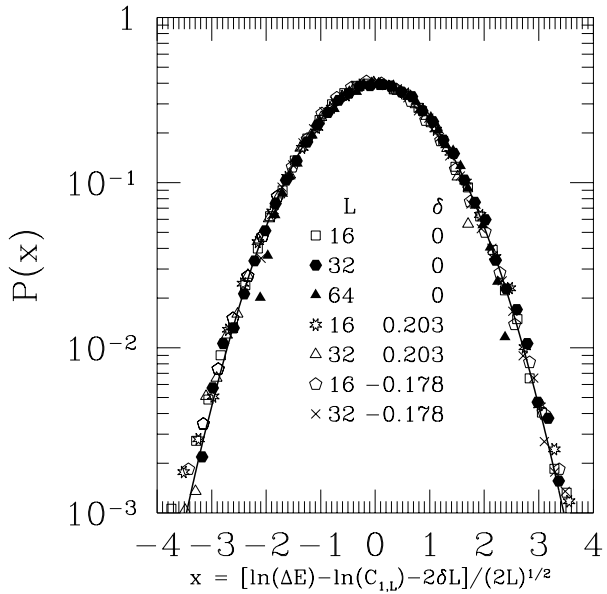}
\caption{
Data for the distribution of $\ln(\Delta E) - \ln(C_{1,l})$, both at and away
from the critical point, $(\delta=0$), showing that it is
a Gaussian with mean and variance given by Eq.~(\protect\ref{L-G}) as expected.
Note that $l_v=1$
%for all $\delta$
for the distributions used here, see 
Eq.~(\protect\ref{lv1}). 
}
\label{gap-cf}
\end{figure}

\subsection{Results Away from Criticality}

\begin{figure}
\epsfxsize=\columnwidth\epsfbox{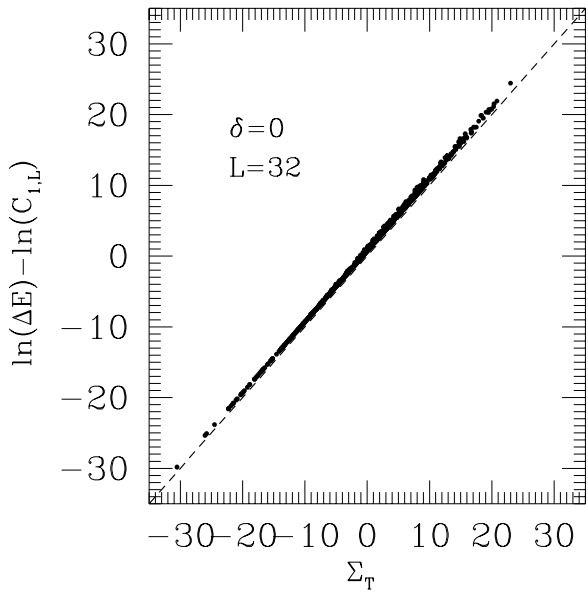}
\caption{
Results for $\ln(\Delta E) - \ln C_({1,L})$ against $\Sigma_T \equiv
\sum_{i=1}^L h_i - \sum_{i=1}^{L-1} J_i$ for $L=32$ at the critical point. Each
point is the result for a single sample and data for 1000 samples are shown.
}
\label{Sigma_T_1_32}
\end{figure}

\begin{figure}
\epsfxsize=\columnwidth\epsfbox{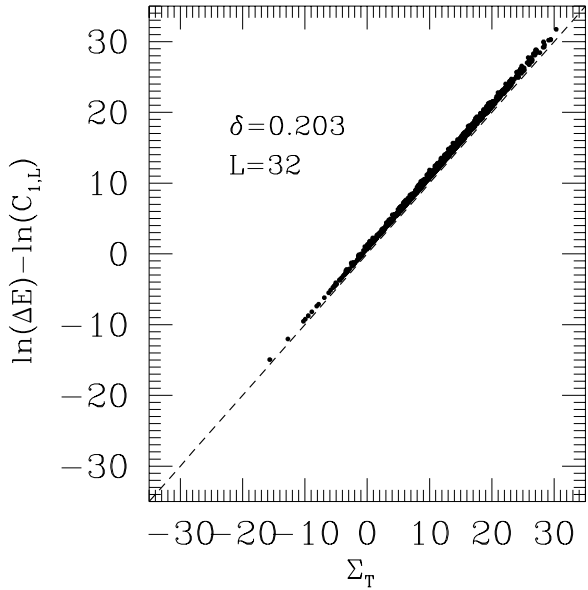}
\caption{
As for Fig.~\protect\ref{Sigma_T_1_32} but for $\delta=0.203$ in the paramagnetic
phase.
}
\label{Sigma_T_1.5_32}
\end{figure}

\begin{figure}
\epsfxsize=\columnwidth\epsfbox{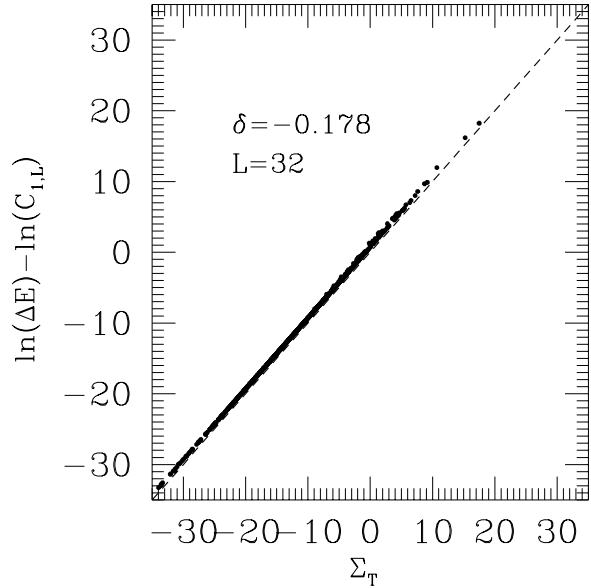}
\caption{
As for Fig.~\protect\ref{Sigma_T_1_32} but for $\delta=-0.178$ in the ferromagnetic
phase.
}
\label{Sigma_T_0.7_32}
\end{figure}

Finally, we have also obtained some results away from the critical point.  A
particularly simple prediction is that $\ln(\Delta E) - \ln C_{1,L}$ should
equal $\Sigma_T \equiv \sum_{i=1}^L \ln h_i - \sum_{i=1}^{L-1}\ln J_i)$,
defined in Eq.~(\ref{eq:L-G}), for each
sample and hence its
distribution should be a Gaussian with mean 
$2\delta L / l_v$ and variance $2 L / l_v$ for {\em all} values of $\delta$,
see Eqs.~(\ref{Sigma_T}) and (\ref{L-G}).
This works very well as shown in Fig.~\ref{gap-cf}, where the solid line is
the predicted Gaussian.  Indeed, we find that, as expected, 
Eq.~(\ref{Sigma_T}) holds for {\em each}
sample. In Figs.~\ref{Sigma_T_1_32}-\ref{Sigma_T_0.7_32} we have plotted
$\ln(\Delta E) - \ln C_{1,L}$ against
$\Sigma_T$
%\equiv \sum_{i=1}^L (\ln h_i - \ln J_i)$,
for $L=32$ and three values of $\delta$. As expected the
data all lie close to the curve $y=x$, indicated by the dashed line.The 
small deviations
presumably come from corrections to the asymptotic form obtained from the RG 
equations.

\section{Conclusions}

In this paper we have studied the probability distributions of two simple
properties of finite length random transverse field Ising chains at and near
the critical point.   In the critical regime, the natural scaling variables are
the logarithms of the gap and the end-to-end correlation function divided by
the square root of the length, $L$.  The distributions of these properties have
been computed explicitly in the scaling limit and the results agree well with
numerical computations.  As noted earlier, \cite{dsf,yr}, the average
correlation function is dominated by rare samples for which the end spins have
correlations of order unity; interestingly, the information about these rare
events is contained, up to non-universal prefactors, in the scaling function of
the distribution of $\ln C_{1,L} /\sqrt L$. 

Although the distribution of the bulk correlation functions are much harder to 
compute analytically, we expect that the same general behavior will obtain for 
them: At the critical point, $\lambda = -\ln C_{0,r}/\sqrt r$ is the natural 
scaling variable and the mean correlations are dominated by a power law 
singularity in the distribution at small $\lambda$ which causes the average 
correlations to decay as $r^{-(3-\sqrt 5)/2}$. Off critical, the typical and 
average correlations both decay exponentially but with the (true) correlation 
length of the average correlations diverging more rapidly than the ``typical 
correlation length" \cite{dsf,yr}. 

This behavior is in striking contrast to that at conventional random critical 
points. If randomness is relevant at a classical random critical 
point, then the natural variable for the distribution of correlations
is $\alpha 
= -\ln C_{0,r}/\ln r$ with \cite{ludwig}
\begin{equation}
d\mbox{Prob}(\alpha) \sim e^{-\ln r~f(\alpha)}d\alpha,
\label{f-alpha}
\end{equation}
a type of distribution sometimes called "multifractal", although in terms of 
the 
log-correlations variable it is quite simple.
The typical log-correlations are thus
$\alpha_{typ} \ln r \pm O(\sqrt{\ln r})$ 
with 
$f(\alpha_{typ})=0$  the minimum value of $f$ ; i.e. they are not very broadly
distributed. Nevertheless, the average correlations are dominated by rare pairs
of spins and decay as a power law with exponent $\alpha_{av} < \alpha_{typ}$.
Off critical, the typical and average correlations decay with different
correlation lengths but the latter is only longer than the former by a constant
factor.

The differences between conventional non-random quantum critical points and the 
random transverse field Ising chain are even more dramatic if one considers the 
decay of correlations in the imaginary time direction.  Instead of the gap 
scaling as a power of the length as at conventional quantum critical points, 
here the logarithm of the gap scales as a
power of the length but with the ratio 
being broadly distributed.  This implies that even the log-correlations in 
imaginary time will be extremely broadly distributed since they will typically 
behave as $-\tau \ln \Delta E_{local}$ with $\Delta E_{local}$ loosely defined 
as
the local gap. 
A numerical study of time-dependent correlations has been given in
Ref.~\onlinecite{y}. It would also be interesting to 
%It would be interesting to
explore these temporal correlations
%which of course really involve a whole spectrum of gaps, in more detail.
in detail analytically.

At this point, it is natural to ask how broad is the class of random quantum
phase transitions for which the behavior is qualitatively similar to that found
here.  In one dimension, the phenomena are very general.  Indeed,
one-dimensional random quantum Potts models \cite{potts} are in exactly the
same universality class for all $q$  and some---but not all---other transitions
in random quantum magnetic chains have many similar features \cite{dsf-af}.  

In
higher dimensions, the situation is less clear.
Simulations on quantum {\em spin glasses} in two and three
dimensions\cite{ry,gbh} have suggested
fairly conventional scaling. However,
recent simulations \cite{py,kr} of
the {\em two}\/-dimensional random quantum Ising {\em ferromagnet}, for which
larger sizes and better statistics can be obtained,
indicate that most of the
features found in one-dimension also apply to this case.
%at zero temperature.
%Various recent simulations
%appear to indicate that the scaling near random quantum ferromagnetic and spin
%glass transitions in two and three dimensions is relatively conventional.
The possibility that some strongly random quantum 
critical points in higher dimensions share some of the exotic features of 
the random transverse field Ising chains is quite intriguing.

\acknowledgments
This work was supported by the National Science Foundation under
grants DMR 9713977, DMR 9630064, DMS 9304580, and via Harvard University's 
MRSEC.

\appendix
\section{}

In this appendix, we briefly consider the distribution of the {\it gap} in long 
chains 
with {\it periodic} boundary conditions like those used in
Ref.~\onlinecite{yr}.
Note that in such systems,
as in bulk systems, the distributions of correlations 
are much more difficult to 
compute analytically than the end-to-end correlations for the free boundary 
case 
and we have no new results on these.

With periodic boundary conditions, it can be shown that, given 
$N$ 
remaining clusters at scale $\Gamma$ in a system of length $L$, the joint 
distribution of the effective couplings and their lengths is simply the 
conditional distribution of $N$ otherwise independent clusters and bonds in an 
infinite system {\it given} that their total length is $L$.  As for the free 
boundary case, this implies correlations between the couplings which can be 
handled by Laplace transform methods as in the main text.  The probability that 
there are $N$ clusters at scale $\Gamma$ can be computed from the RG equation 
that couples the flow of the $N$-cluster distribution and that of $N-1$ 
clusters 
when one of the $N$ bonds (or clusters) is decimated. As in the free case, $G$, 
minus the 
logarithm of the gap, is 
determined by the scale at which the last cluster is decimated. We obtain
\begin{eqnarray}
d\mbox{Prob}(G) & = & L~ dG~\raisebox{-1.8ex}{
$\stackrel{\textstyle LT^{-1}}{L\to y}$} \Bigl\{\frac{\Delta ^2}{\sinh 
^2(G\Delta )\left[\Delta \coth (G\Delta) - \delta \right]} \nonumber \\
&+&
\int_0^G d\Gamma \frac{\Delta^2}{\sinh ^2(\Gamma \Delta)}e^{-(G-\Gamma)[\Delta 
\coth(\Gamma\Delta)+\delta]}\Bigr\}.
\label{per-gap}
\end{eqnarray}
We have not been able to perform the integrations in Eq.~(\ref{per-gap}) 
analytically, but the asymptotic behavior of the scaling distribution at the 
critical point can be found.  In terms of the rescaled variable $g=G/\sqrt{L}$,
we find for small $g$ 
\begin{equation}
d\mbox{Prob}(g) \approx dg ~ \frac{\pi^2}{2g^3}e^{-\pi^2/(4g^2)} ,
\end{equation}
very similar to the free-end case Eq.~(\ref{eq:A53}).  For large 
$g$,
\begin{equation}
d\mbox{Prob}(g) \approx dg~ \frac{g}{2}e^{-\frac{1}{4}g^2}
\end{equation}
again quite similar to the free-end result Eq.~(\ref{eq:A53}).  These limiting 
forms are a reasonable fit to the numerical data of Ref.~\onlinecite{yr}.


\begin{references}
%\bibitem{th}
%M.~J.~Thill and D.~A.~Huse, Physica A, {\bf 15}, 321 (1995).

\bibitem{dsf}
D.~S.~Fisher, Phys. Rev. Lett. {\bf 69}, 534 (1992); Phys. Rev. B {\bf
51}, 6411 (1995).

%\bibitem{mccoy}
%B.~M.~McCoy, Phys. Rev. Lett. {\bf 23}, 383 (1969); Phys. Rev. {\bf
%188}, 1014 (1969).
%
%\bibitem{griffiths}
%R.~B.~Griffiths, Phys. Rev. Lett. {\bf 23}. 17 (1969).
%
%%\bibitem{essen}
%%A.~B.~Harris, Phys. Rev. B {\bf 12}, 203 (1975).
%%
%\bibitem{mw}
%B.~M.~McCoy and T.~T.~Wu, Phys. Rev. {\bf 176}, 631 (1968); {\bf 188},
%982 (1969).

\bibitem{yr}
A.~P.~Young and H.~Rieger, Phys. Rev. B, {\bf 53}, 8486 (1996).

\bibitem{IR} F. Igl\'oi and H.~Rieger, cond-mat 9709260, have also
recently
studied end-to-end spin correlations

\bibitem{sm}
R.~Shankar and G.~Murphy, Phys. Rev. B, {\bf 36}, 536 (1987).

%\bibitem{lsm}
%E.~Lieb, T. Schultz and D.~Mattis, Ann. Phys. (NY) {\bf 16}, 407 (1961).
%
%\bibitem{pfeuty}
%P.~Pfeuty, Ann. Phys. (NY) {\bf 27}, 79 (1970); Th\`ese, Universit\'e de
%Paris, (1970).
%
%\bibitem{katsura}
%S.~Katsura, Phys. Rev. {\bf 127}, 1508 (1962).

%\bibitem{ri}
%H.~Rieger and F.~Igloi, cond-mat/9704152.
%
%\bibitem{sy}
%S.~Sachdev and A.~P.~Young, Phys. Rev. Lett. {\bf 78}, 2220  (1997).

\bibitem{dprob}
We use notations like $d\mbox{Prob}(x|A)$ to mean the probability that the
variable $x$ is in the interval $(x,x+dx)$ given the condition $A$.

\bibitem{y}
A.~P.~Young, Phys. Rev. B {\bf 56}, 11691 (1997).

\bibitem{snm}
J.~Stolze, A.~N\"oppert and G. M\"uller, Phys. Rev.  {\bf 52},
4319 (1995); H.~Asakawa, Physica A {\bf 233}, 39 (1996).
\bibitem{ludwig} A.A.W. Ludwig, Nucl. Phys. {\bf B330} 639 (1990).

\bibitem{dsf-af}
D.S. Fisher, Phys. Rev. B50, 3799 (1994).

\bibitem{potts}
T. Senthil and S.N. Majumdar,
Phys. Rev. Lett. {\bf 76}, 3001 (1996).

\bibitem{ry}
H.~Rieger and A.~P.~Young, Phys. Rev. Lett. {\bf 72}, 4141 (1994).

\bibitem{gbh}
M.~Guo, R.~N.~Bhatt and D.~A.~Huse, Phys. Rev. Lett. {\bf 72}, 4137
(1994).

\bibitem{py}
C. Pich and A. P. Young cond-mat/9802108.

\bibitem{kr}
H. Rieger and N. Kawashima cond-mat/9802104.

\end{references}
\end{document}